\documentclass[aps,prl,twocolumn,showpacs,groupedaddress]{revtex4-1}
\usepackage{graphicx}
\begin{document}
\title{Thermodynamic behavior of supercritical matter}
\author{Dima Bolmatov$^{1}$}
\thanks{d.bolmatov@gmail.com}
\author{V. V. Brazhkin$^{2}$}
\author{K. Trachenko$^{1,3}$}
\address{$^1$ School of Physics and Astronomy, Queen Mary University of London, Mile End Road, London, E1 4NS, UK}
\address{$^2$ Institute for High Pressure Physics, RAS, 142190, Moscow, Russia}
\address{$^3$ South East Physics Network}
\begin{abstract}
Since their discovery in 1822, supercritical fluids have been of enduring interest, and have started to be deployed in many important applications. Theoretical understanding of the supercritical state is lacking, and is seen to limit further industrial deployment. Here, we study thermodynamic properties of the supercritical state, and discover that specific heat shows a crossover between two different regimes, an unexpected result in view of currently perceived homogeneity of supercritical state in terms of physical properties. We subsequently formulate a theory of system thermodynamics above the crossover, and find good agreement between calculated and experimental specific heat with no free fitting parameters. We derive a power law and analyze supercritical scaling exponents in the system above the Frenkel line. In this theory, energy and heat capacity are governed by the minimal length of the longitudinal mode in the system only, and do not explicitly depend on system-specific structure and interactions.
\end{abstract}

\maketitle
Statistical mechanics is the art of predicting the behavior of a system with a large number of degrees of freedom, given the laws governing its microscopic behavior. The statistical description of liquids, in comparison with the solid and gas phases, is incomplete. The problem of formulating a rigorous mathematical description of liquids has always been regarded as much more difficult than that of the kinetic theory of gases or solid-state theory, stimulating the ongoing research  \cite{angell,kivel,harow,wolyn,birodi,dyre,larini,pathintegral,visco}. Due to the simultaneous presence of strong interactions and large atomic displacements, common models and approximations used for gases and solids, do not apply to liquids. For this reason, liquids do not generally fall into any simple classification, and have been mostly treated as general many-body systems as a result.

In recent years, a significant effort has been devoted to investigation of various properties of supercritical fluids \cite{jon,deb,kesse,book1,book2}. This has been an exciting field with a long history since 1822 when Baron Charles Cagniard de la Tour discovered supercritical fluids while conducting experiments involving the discontinuities of the sound in a sealed cannon barrel filled with various fluids at high temperature \cite{bertran}. More recently, supercritical fluids have started to be deployed in several important applications, ranging from the extraction of floral fragrance from flowers to applications in food science such as creating decaffeinated coffee, functional food ingredients, pharmaceuticals, cosmetics, polymers, powders, bio- and functional materials, nano-systems, natural products, biotechnology, fossil and bio-fuels, microelectronics, energy and environment \cite{book1,book2,bruner}. Much of the excitement and interest of the past decade is due to the enormous progress made in increasing the power of relevant experimental tools \cite{simone,pai,shah,ruocco}. The development of new experimental methods and improvement of existing ones continues to play a important role in this field \cite{monaco,tokmakof,brad,george,stanley}, with recent research focusing on dynamic properties of fluids \cite{frad,bert,mason,tanaka,indu,flen}.

High density and high thermal motion are two main properties responsible for efficient cleaning, dissolving and extracting abilities of supercritical fluids in the above industrial applications. From the point of view of practical applications, supercritical fluids have got the best of both worlds: high density comparable to ordinary liquids and solids and high thermal motion and diffusivity approaching that of gases. Notably, it is this very combination that presents a formidable problem to the theory: high density and strong interactions mean that theories and approximations used for dilute gases do not apply \cite{gases}. Early Enskog and related kinetic approaches to gases were followed by more extensive developments, yet they do not adequately describe dense systems with strong interactions and many-body correlations \cite{gases} such as supercritical fluids. One general issue with extending gas-like approaches to fluids was noted earlier: in a system with strong interactions, the system energy strongly depends on the type of interactions, and is therefore system-specific, ruling out the possibility to develop a theory universally applicable to many fluids, in contrast to gases and solids \cite{landau}.

In addition to theoretical challenges, the lack of fundamental understanding is seen as an obstacle towards wider deployment of supercritical fluids in industrial applications, primarily due to the absence of guidance regarding pressure and temperature at which the desired properties are optimized as well as the possibility to use new systems \cite{book1}.

In this paper, we focus on the thermodynamic properties of the supercritical state. On the basis of molecular dynamics simulations, we find that specific heat shows a crossover between two different dynamic regimes of the low-temperature rigid liquid and high-temperature non-rigid gas-like fluid. The crossover challenges the currently held belief that no difference can be made between a gas and a liquid above the critical point and that the supercritical state is homogeneous in terms of physical properties \cite{hansen}. We subsequently formulate a theory of system thermodynamics and heat capacity above the crossover. In this theory, energy and heat capacity are governed by the minimal length of the longitudinal mode in the system only, and do not depend on system-specific structure and interactions. We further study the predicted relationship between supercritical exponents of heat capacity and viscosity. A good agreement is demonstrated between calculated and experimental data for noble and molecular supercritical fluids with no free fitting parameters.

\section{Results}

\subsection{Dynamic crossover of the specific heat}

We start with molecular dynamics simulations of a model liquid. Our primary aim here is to show that specific heat, $c_V$, shows a crossover in the supercritical region of the phase diagram. This result is unexpected in view of currently perceived homogeneity of supercritical state in terms of physical properties.

\begin{figure}
	\centering
\includegraphics[scale=0.6]{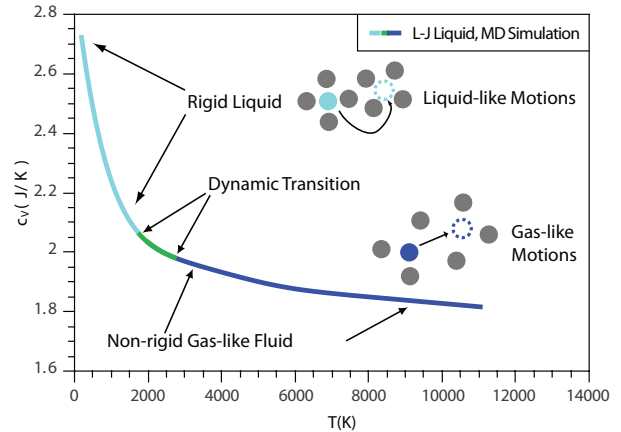}
\caption{ {\bf Heat capacity of binary Lennard-Jones fluid}. Calculated $c_V$ showing the crossover and continuous dynamical transition around $c_V\approx 2$, ($k_{\rm B}=1$). The crossover takes place between different dynamical regimes of the rigid liquid and non-rigid supercritical fluid.}
	\label{fig1}
\end{figure}

Using molecular dynamics simulations (see {\bf Methods}), we have simulated the binary Lennard-Jones (LJ) fluid. We have simulated the system with 64000 atoms using constant-volume (nve) ensemble in the wide temperature range (see Fig.\ref{fig1}) well extending into the supercritical region. Indeed, the temperature range in Figure 1 is between about 2T$_{\rm c}$ and 70T$_{\rm c}$, where T$_{\rm c}$ is the critical temperature of Ar, T$_{\rm c}\approx$ 150 K; the simulated density, 2072 kg m$^{-3}$, corresponds to approximately four times the critical density of Ar. From the energy of the system $E$ at each temperature, we calculate constant-volume specific heat, $c_V$, as $c_V=\frac{1}{N}\frac{dE}{dT}$ ($k_{\rm B}=1$).

We observe that $c_V$ decreases steeply from the solid-state value of about 3$k_{\rm B}$ at low temperature to approximately 2$k_{\rm B}$ around 2000 K. The steep decrease is followed by crossing over to a considerably weaker temperature dependence. This crossover is a new effect not reported in previous MD simulations. We further observe that the crossover takes place around $c_V\approx 2 k_{\rm B}$. This value of $c_V$, $c_V=2 k_{\rm B}$ is non-coincidental, and corresponds to the crossover taking place across the Frenkel line \cite{brazhkin0,brazhkin} as discussed below.

Crossing the Frenkel line corresponds to the qualitative change of atomic dynamics in a liquid. In liquids, atomic motion has two components: a solid-like, quasi-harmonic vibrational motion about equilibrium locations and diffusive gas-like jumps between neighboring equilibrium positions. As the temperature increases, a particle spends less time vibrating and more time diffusing. Eventually, the solid-like oscillating component of motion disappears; all that remains is the gas-like ballistic motion. That disappearance, a qualitative change in particle dynamics, corresponds to crossing the Frenkel line, the transition of the substance from the liquid dynamics to the gas dynamics. This transition takes place when liquid relaxation time $\tau$ ($\tau$ is liquid relaxation time, the average time between consecutive atomic jumps at one point in space \cite{frenkel}) approaches its minimal value, $\tau_{\rm D}$, the Debye vibration period. As recently discussed \cite{brazhkin0,brazhkin}, crossing the Frenkel line is accompanied by qualitative changes of most important properties of the system, including diffusion, viscosity, thermal conductivity and dispersion curves \cite{brazhkin0,brazhkin}. One of these properties is the crossover of the speed of sound and appearance of the high-frequency "fast" sound \cite{ruocco,psd1,psd2,psd3}. At the microscopic level, the Frenkel line can be identified by the loss of oscillations in the velocity autocorrelation function \cite{brazhkin}, a point studied in detail elsewhere.

Fig. 1 implies that in addition to the dynamic properties discussed above, the thermodynamics of the system changes at the Frenkel line too. This is an important general insight regarding the behavior of the supercritical state.

The initial steep decrease of $c_V$ from about $3 k_{\rm B}$ to $2 k_{\rm B}$ can be quantitatively explained by the progressive loss of two transverse waves with frequency $\omega>1/\tau$ \cite{bolmatov,tg,trachenko}. Physically, this picture is based on Frenkel's prediction that on time scale shorter than $\tau$, liquid is a solid, and therefore supports rigidity and solid-like transverse waves at short times, or at frequency larger than $\omega>1/\tau$ \cite{helium}. When $\tau$ approaches its maximal value, $\tau_{\rm D}$, the liquid can not sustain transverse waves at any frequency \cite{pilgrim}. Consequently, the potential energy of the system is due to the longitudinal mode only, giving the total energy of $2NT$ and specific heat of 2$k_{\rm B}$ \cite{bolmatov,trachenko}. Hence, the decrease of $c_V$ from about $3 k_{\rm B}$ to $2 k_{\rm B}$ corresponds to the region of a "rigid" liquid where short-time solid-like rigidity and high-frequency transverse waves exist. On the other hand, the liquid is unable to sustain transverse waves at any available frequency above the Frenkel line. Instead, the liquid enters a new dynamic "non-rigid" gas-like regime where oscillatory component of particles is lost and the motion becomes purely collisional as in a gas.

We therefore need to develop a new theory capable of describing thermodynamics of supercritical matter above the Frenkel line where the system enters the new dynamic regime and where $c_V$ falls below $2 k_{\rm B}$ and approaches the ideal-gas value of $3/2 k_{\rm B}$ at high temperature (see Fig. \ref{fig1}).

\subsection{Thermodynamic theory of supercritical state}

We now focus on the theory of the non-rigid gas-like liquid above the Frenkel line, and add a new proposal regarding how the system energy can be evaluated. As temperature rises in the ballistic gas-like regime and kinetic energy increases, the mean free path $l$, the average distance between particle collisions, increases. At the Frenkel line where the ballistic regime starts, $l$ is comparable to interatomic separation. In the limit of high temperature where the particle's kinetic energy is much larger than potential energy, $l$ tends to infinity as in the non-interacting ideal gas. Our proposal is that $l$ determines the shortest wavelength of the longitudinal mode that exists in the system, $\lambda$, because below this length the motion is purely ballistic and therefore can not be oscillatory: $\lambda=l$. On the other hand, the longitudinal modes with larger wavelength are supported, and represent the excitations existing in the supercritical system.

We note that the existence of long-wavelength longitudinal waves in a gas, sound, is well-known. What is new here is that we propose that the contribution of the longitudinal waves to the energy of the gas-like supercritical system starts from very short wavelengths comparable with interatomic separation $a$. In this sense, we are extending the solid state concepts (e.g. short-wavelength solid-like phonons with Debye density of states, see below) to the new area of gas-like supercritical state where these ideas have not been hitherto contemplated. Indeed, it is well established experimentally that dynamics in sub-critical liquids shows solid-like character, in that liquids can sustain high-frequency propagating modes down to wavelengths on the atomic scale, with solid-like dispersion relations \cite{pilgrim}. Importantly, recent experimental evidence shows that the same applies to supercritical fluids \cite{ruocco,jcp}.

Here and elsewhere, our discussion of liquid vibrational states includes an important point. Namely, a disordered system such glass or liquid supports non-decaying collective excitations obtainable from the secular equation involving the force matrix constructed from the amorphous glass or liquid structure. Harmonic plane waves naturally decay in liquids as in any non-crystalline systems, yet importantly they are clearly seen in fluids experimentally as quasi-linear solid-like dispersion relations even in low-viscous liquids \cite{pilgrim,monaco,hosoka}, leading to the quadratic density of states $g(\omega)\propto\omega^2$. A detailed discussion of this point is forthcoming.

In the proposed theory, the energy of the non-rigid supercritical fluid per particle includes the contribution from the kinetic energy, $K=\frac{3}{2}k_{\rm B} T$ and the potential energy of the longitudinal phonons with wavelengths larger than $l$. Using the equipartition theorem $\langle P_l\rangle=\frac{\langle E_{l}\rangle}{2}$, where $\langle E_{l}\rangle$ is the energy of the longitudinal phonons, we write

\begin{equation}
E=\frac{3}{2}k_{\rm B}T+\frac{1}{2}\int\limits_{0}^{\omega_{0}} \varepsilon(\omega,T)g(\omega)d\omega+E_{anh}
\label{energy}
\end{equation}

\noindent where the upper integration limit $\omega_0$ is given by the shortest wavelength in the system, $\lambda$: $\omega_0=\frac{2\pi}{\lambda}c$, $c$ is the speed of sound, $\varepsilon(\omega,T)$ is the mean energy of harmonic oscillator, $\varepsilon(\omega,T)=\frac{\hbar\omega}{2}+\frac{\hbar\omega}{e^{\hbar\omega/T}-1}$, or $\varepsilon(T)=k_{\rm B}T$ in the classical case, and $E_{anh}$ is the anharmonic contribution to the phonon energy.

The second term in Eq. (\ref{energy}) can be calculated using the Debye density of states, $g(\omega)\equiv\frac{3}{\omega_{\rm D}^{3}}\omega^2$, giving $k_{\rm B}T\frac{\omega^3}{\omega_{\rm D}^3}$, or $k_{\rm B}T\frac{a^3}{\lambda^3}$, where $a$ is the interatomic separation. The use of the quadratic density of states is supported by the experimental evidence showing solid-like quasi-linear dispersion relationships in supercritical fluids \cite{ruocco, jcp} similar to the subcritical liquids. $E_{anh}$ can be evaluated in the Gr\"{u}neisen approximation from the softening of phonon frequencies with temperature, with the result that the energy is modified as $E\rightarrow E\left(1+\frac{\alpha T}{2}\right)$, where $\alpha$ is the coefficient of thermal expansion \cite{tg,trachenko}. Then, the energy of non-rigid gas-like supercritical fluid becomes:

\begin{eqnarray}\label{energy}
E=\frac{3}{2}k_{\rm B}T+\left(1+\frac{1}{2}\alpha T\right)\frac{1}{2}k_{\rm B}T\frac{a^{3}}{\lambda^{3}}
\label{energy1}
\end{eqnarray}

We observe that when $\lambda\approx a$ at the Frenkel line, Eq. ({\ref{energy1}) gives $E=2 k_{\rm B}T$ and the specific heat of $2$ (here we neglect the small term $\propto \alpha T$). This corresponds to the crossover of $c_V$ in Figure 1 to the gas-like regime as discussed above. When $\lambda$ increases on temperature increase in the gas-like regime, Eq. ({\ref{energy1}) predicts that $c_V$ tends to $\frac{3}{2}$, the ideal-gas value as expected.

\subsection{Comparison with experimental data}

As discussed above, $\lambda$ is given by the particle mean free path in the non-rigid gas-like regime, and can therefore be calculated from $\eta=\frac{1}{3}\rho\bar{u}\lambda$, where $\eta$ is viscosity, $\rho$ is density and $\bar{u}$ is average velocity. Therefore, Eq. (\ref{energy1}) provides an important relationship between the energy of the non-rigid supercritical liquid and its gas-like viscosity. We now check this relationship experimentally, by comparing the specific heat $c_{V}=\frac{dE}{dT}$ predicted by Eq. (\ref{energy1}) and the experimental $c_V$.

We have used the National Institute of Standards and Technology (NIST Chemistry WebBook, http://webbook.nist.gov/chemistry/fluid) database. We aimed to check our theoretical predictions and selected the isochoric data of several supercritical noble and molecular liquids in a wide range of temperature and density. We note that $\lambda$ in Eq. (\ref{energy}) changes in response to both temperature and density: $\lambda$ increases with temperature and decreases with density. Practically, the range of isochoric data in the NIST database is fairly narrow in terms of density compared to the range of temperature. We therefore analyze the temperature behavior of $c_v$ and $\eta$ along several isochores. Our choice of liquids is dictated by the availability of isochoric data in the supercritical region. For each density, we fit experimental viscosity to $\eta=A_{0}+A_{1} T^{A_{2}}$, calculate $\lambda$ from $\eta=\frac{1}{3}\rho\bar{u}\lambda$ and subsequently use $\lambda$ in Eq. (\ref{energy1}) to calculate $c_V$.

We note that Debye model is not a good approximation in molecular liquids where the frequency of intra-molecular vibrations considerably exceeds the rest of frequencies
in the system (3340 K in N$_2$ and 3572 K in CO). However, the intra-molecular modes are not excited in the temperature range of experimental $c_V$ (see Figures 3--4). Therefore, the contribution of intra-molecular motion to cv is purely rotational, $c^{rot}$. On the other hand, the rotational motion is excited in the considered temperature range, and is therefore classical, giving $c^{rot}=R$ for linear molecules in N$_2$ and CO. Consequently, $c_V$ for molecular liquids shown in Figure 4 correspond to heat capacities per molecule, with $c^{rot}$ subtracted from the experimental data.

We also note that experimental isochoric $c_V$ is affected by lambda-like critical anomalies (see Figs. 2, 3) because the isochoric NIST data do not extend far above the critical point. Here, we do not consider critical effects related to phase transitions, and therefore fit the data at temperatures that are high enough to be affected by the lambda-anomaly at the phase transition. In Figures 2 and 3 we observe good agreement between experimental and predicted $c_V$, in view of (a) 2-5\% uncertainty of experimental $c_V$ and $\eta$ (see NIST Chemistry WebBook), (b) approximations introduced by the Debye model and (c) increased curvature of $c_V$ at low temperature due to proximity of lambda-anomalies that are not taken into account by the theory. Notably, the agreement is achieved without using free fitting parameters because $\rho$, $\alpha$, $\bar{u}$ and $a$ are fixed by system properties. Values of these parameters used in Eq.(\ref{energy}) are in good agreement with their experimental values.

\begin{figure}
	\centering
\includegraphics[scale=0.6]{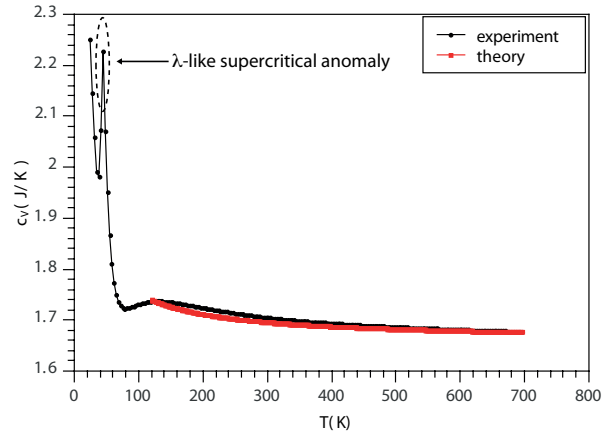}
\caption{{\bf Heat capacity of liquid Ne}. Experimental and calculated $c_{V}$ ($k_{\rm B}=1$) for non-rigid supercritical fluid Neon. Experimental $c_{v}$ and $\eta$ are taken from the NIST database at density $700$ kg m$^{-3}$. Value of $\alpha$ used in the calculation is $3.5\times10^{-3}$ K$^{-1}$.}
	\label{fig2}
\end{figure}

\begin{figure*}
	\centering
\includegraphics[scale=0.78]{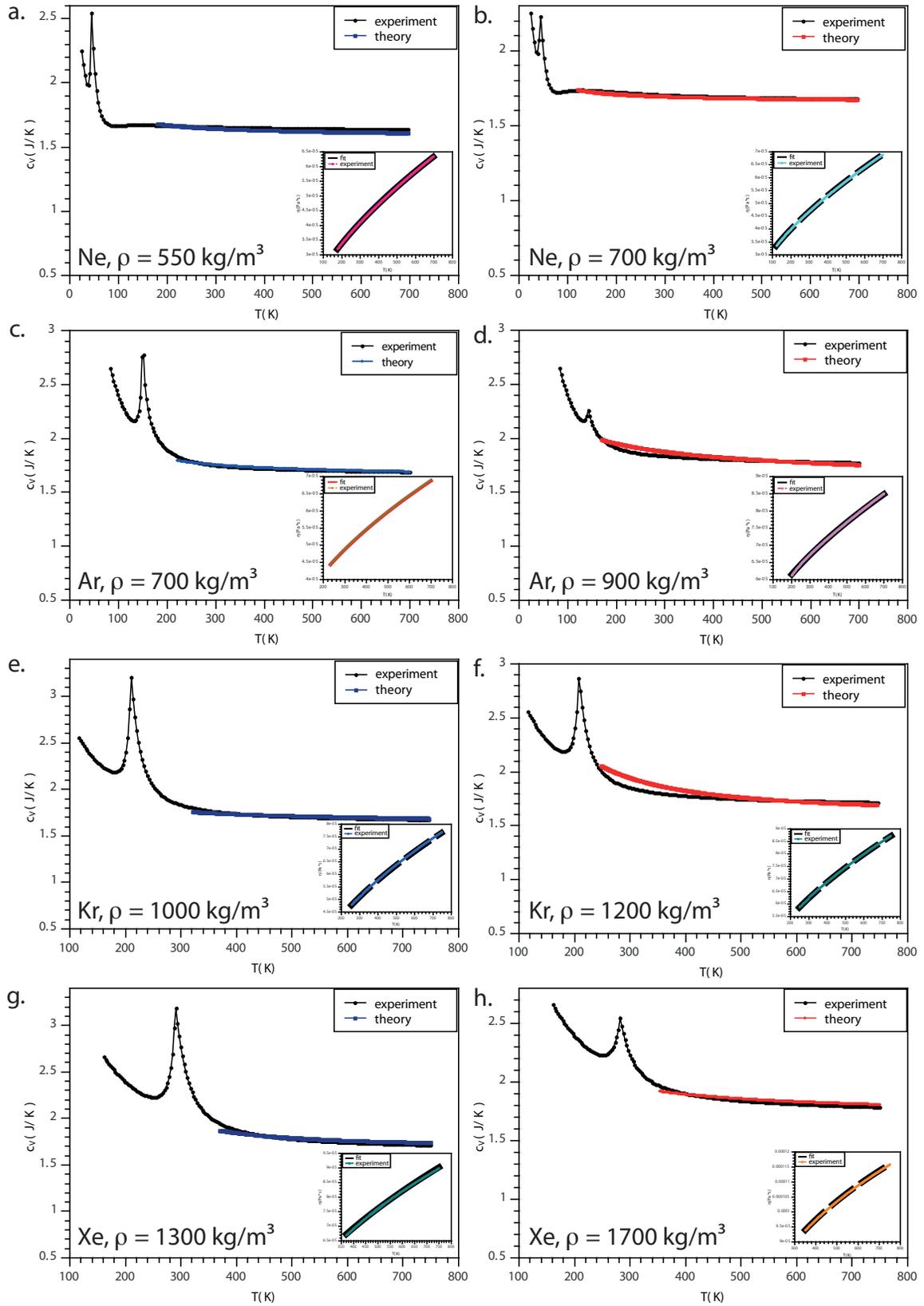}
\caption{{\bf Heat capacity of noble gas liquids}. Experimental and calculated $c_{V}$ ($k_{\rm B}=1$) for noble non-rigid supercritical fluids. Experimental $c_{v}$ and $\eta$ are taken from the NIST database  at different densities as shown in the Figure. Values of $\alpha$ used in the calculation are $3.5\times10^{-3}$ $K^{-1}$ (Ne), $3.3\times10^{-3}$ $K^{-1}$ (Ar), $1.8\times10^{-3}$ K$^{-1}$ (Kr) and $8.2\times10^{-4}$ K$^{-1}$ (Xe). The uncertainty of both experimental heat capacities and viscosities is about 2-5\%. Insets show viscosity fits.}
	\label{fig3}
\end{figure*}

\begin{figure*}
	\centering
\includegraphics[scale=0.6]{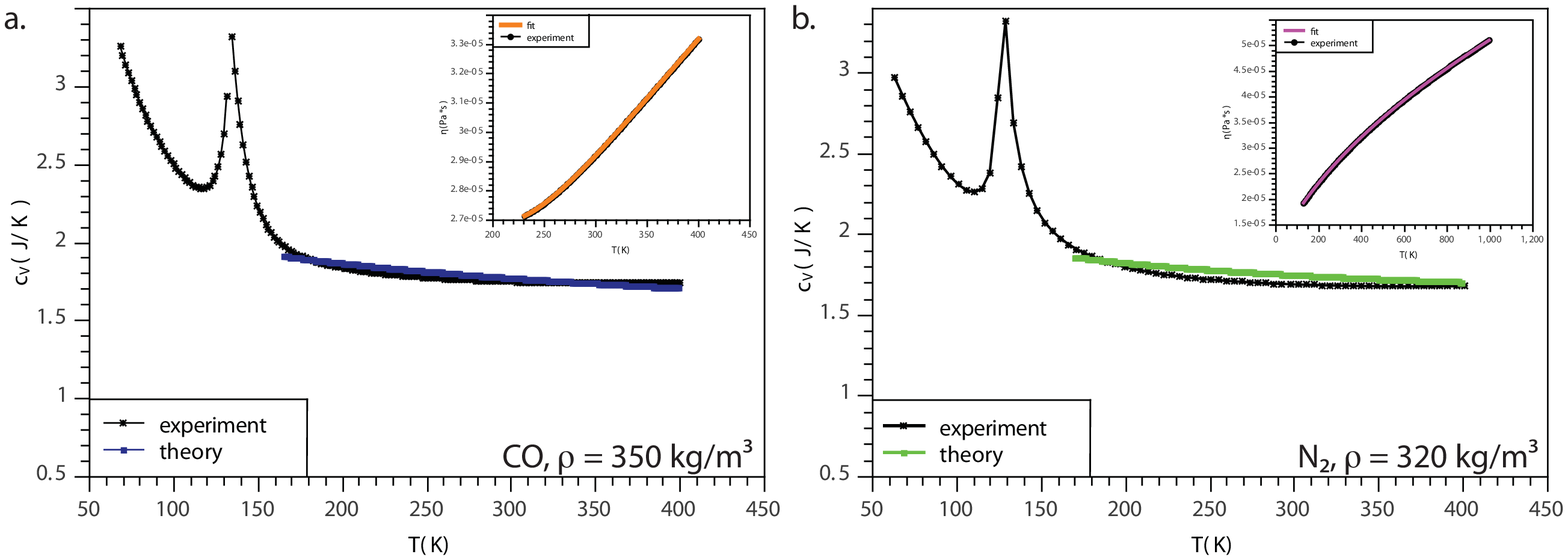}
\caption{{\bf Heat capacity of molecular liquids}. Experimental and calculated $c_{V}$ ($k_{\rm B}=1$) for molecular non-rigid supercritical fluids. Experimental $c_{v}$ and $\eta$ are taken from the NIST database. Values of $\alpha$ used in the calculation are $6.5\times10^{-3}$ K$^{-1}$ (CO) and $8.5\times 10^{-3}$ K$^{-1}$ (N$_2$). The uncertainty of both experimental heat capacities and viscosities is about 2-5\%. Insets show viscosity fits.}
	\label{fig4}
\end{figure*}

Another way to compare our theory and experimental data is to study supercritical exponents of specific heat and viscosity. Indeed, if $\eta$ in the non-rigid gas-like supercritical region can be approximated as a power law, $\eta\propto T^{\gamma}$, then Eq. (\ref{energy1}) makes two predictions. First, $E$ and $c_V$ should also follow power laws. Second, Eq. (\ref{energy1}) provides a specific relationship between the scaling exponents of $\eta$ on one hand and $E$ and $c_V$ on the other hand, the relationship that we check below.

Apart from comparing theory and experiments, studying the supercritical scaling exponents is interesting in the wider context of scaling behavior of physical properties. In the area of phase transitions, the scaling behavior idea has been a crucial element in the subject of liquids and other systems from the nineteenth century onwards \cite{stan}. Critical points occur in a great variety of systems \cite{mishima,subir,chandler,watanabe}, yet there is a considerable degree of similarity in the way in which systems approach the critical point \cite{anisimov}. Here, the calculation of critical exponents serves as one of the main test of the theories.

Once analyzed within our theory below, experimental temperature dependence of $\eta$, $\bar{u}$, $E$ and $c_{V}$ (see NIST Chemistry WebBook) can be fairly well approximated by the power law above the Frenkel line.  Therefore, we use the following power-law relationships:

\begin{eqnarray}\label{hc}
\eta &=& const\times T^{\gamma} \\
\bar{u} &=& const\times T^{\frac{1}{2}} \\
E &=& 1.5\times T+const\times T^{\delta} \\
c_{V} &=& 1.5+\frac{const}{T^{\xi}} \label{cvt}
\end{eqnarray}

From $\eta=\frac{1}{3}\rho\bar{u}\lambda$ ($k_{\rm B}=1$), $\lambda=const\times T^{\gamma-0.5}$. Using $\lambda$ in Eq. (\ref{energy1}) gives for the energy (neglecting the small $\propto\alpha T$ term):
\begin{eqnarray}\label{en}
E &\propto & T^{2.5-3\gamma}
\end{eqnarray}
and specific heat:
\begin{eqnarray}\label{hc1}
c_{V} &\propto & (2.5-3\gamma)T^{1.5-3\gamma}
\end{eqnarray}

Then, from Eqs. (\ref{hc}), (\ref{en}) and (\ref{hc1}) we find the following relationships between the power-law scaling exponents:

\begin{equation}\label{enlaw}
2.5-3\gamma=\delta
\end{equation}

\begin{equation}\label{hclaw}
3\gamma-1.5=\xi
\end{equation}

We note that Eqs. (\ref{en}) and (\ref{hc1}), together with the general requirement for $c_V$ to be positive and the experimental requirement for $c_V$ to decrease with temperature, imply that $\gamma$ should be in the range $\frac{1}{2}<\gamma<\frac{5}{6}$. This is the case for the experimental data we examined, as will be shown below. We also note that $\gamma=\frac{1}{2}$ corresponds to the non-interacting ideal gas and $\eta\propto T^{0.5}$.

We now check the Eq. (\ref{hclaw}) experimentally. Using the NIST database and Eq.(\ref{cvt}), we calculate $\xi$ from the slope of $\log{\left(c_V-1.5\right)}$ vs. $\log{\left(\frac{1}{T-T_{\rm F}}\right)}$, where $T_{\rm F}$ corresponds to the temperature on the Frenkel line and $1.5$ is the asymptotic gas value of specific heat. Using Eq. (\ref{hclaw}), we calculate the predicted theoretical value $\gamma_{th}$ and compare it with the experimental $\gamma_{exp}$ obtained from fitting the experimental viscosity to $\eta\propto T^{\gamma_{exp}}$.

\begin{table}[ht]
\centering
\begin{tabular}{c c c c c}
\hline\hline
Supercritical Fluid  & $<\xi>$ & $<\gamma_{th}>$ & $<\gamma_{exp}>$ \\ [0.5ex]
\hline
Ar & 0.17 & 0.56 & 0.55\\
Ne & 0.17 & 0.56 & 0.64 \\
Kr & 0.18 & 0.56 & 0.60 \\
Xe & 0.18 & 0.56 & 0.55 \\
CO & 0.17 & 0.56 & 0.61 \\
N$_2$ & 0.27 & 0.59 & 0.59 \\ [1ex]
\hline
\end{tabular}
\caption{Power-law exponents $\xi$, $\gamma_{exp}$ and $\gamma_{th}$, averaged over different densities: Ne(600 kg m$^{-3}$, 700 kg m$^{-3}$, 900 kg m$^{-3}$), Ar (700 kg m$^{-3}$, 800 kg m$^{-3}$, 900 kg m$^{-3}$), Kr(1100 kg m$^{-3}$, 1300 kg m$^{-3}$, 1500 kg m$^{-3}$), Xe(1300 kg m$^{-3}$, 1500 kg m$^{-3}$, 1700 kg m$^{-3}$), CO(350 kg m$^{-3}$) and N$_{2}$(320 kg m$^{-3}$).}
\label{table1}
\end{table}

In Table \ref{table1} we show $\xi$, $\gamma_{exp}$ and $\gamma_{th}$, averaged over several data sets taken along the isochores at several different densities. We observe the overall good agreement between $\gamma_{exp}$ and $\gamma_{th}$. We further observe that the power exponent of specific heat, $\xi$, is close to $0.2$ for different systems. In our theory, the similarity of temperature behavior of $c_V$ is due to temperature dependence of $\lambda$. This point is discussed in the next section in more detail.

\section{Discussion}

Our first important observation in this work is that contrary to the current belief, the thermodynamic properties of the supercritical state are not homogeneous. Instead, the specific heat shows a crossover related to the change of particle dynamics, which we attributed to the recently introduced Frenkel line.

We have subsequently focused on thermodynamic properties of supercritical fluids above the Frenkel line. Here, we faced the problem of strong interactions, the long-persisting challenge in condensed matter physics. Indeed, strong interactions imply that approximations used for dilute gases do not apply to real dense liquids \cite{gases}. If we consider realistic strong interactions (assuming that interactions are known and can be represented in analytical form, the assumption that is valid for a relatively small number of simple systems only) and structural correlations that often include those beyond two-body correlations, we quickly find that the problem becomes intractable. Further, strong interactions, coupled with their specificity in different systems, have been suggested to preclude the calculation of energy and heat capacity in general form from the outset \cite{landau}.

In this paper, we addressed the problem in a different way, by substituting all potentially complicated effects of interactions and structural correlations by one physical quantity, the minimal wavelength of the longitudinal mode in the system $\lambda$. This has enabled us to rationalize the experimental behavior of $c_V$ as well as provide the relationship between different physical properties and experimental outcomes (e.g. relationship between $c_V$ and $\eta$). Notably, our approach unveils similarity of thermodynamics of supercritical state in the following sense. First, $c_V$ does not explicitly depend on system details such as structure and interactions but on $\lambda$ only. Fluids may have very different structure and interactions, yet our theory predicts the similarity of their thermodynamic behavior as long as $\lambda$ behaves similarly in those systems. Second, and more specifically, our approach predicts that supercritical scaling of thermodynamic properties such as heat capacity is governed by viscosity scaling. Consequently, similar temperature scaling of viscosity gives similar temperature scaling of thermodynamic properties. We note here that we have mostly dealt with systems with fairly simple interatomic interactions whereas the found similarity of thermodynamic behavior may not hold in systems with the hierarchy of interactions and non-trivial structural transformations such as water \cite{water}.

\section{Methods}
We have used DL-POLY molecular dynamics simulation code \cite{todorov} to run the system with 64000 atoms in the constant-volume (nve) ensemble at different temperatures. We have used 3000  processors of the high-throughput cluster to simulate 500 temperature points in the temperature range of about 0--12000 K shown in Figure 1. We have used the model Lennard-Jones potential \cite{kobLJ} to simulate model liquids. Each structure was equilibrated for 25 ps, and the system energy and other properties were averaged for the following 25 ps of the production run. The timestep of 0.001 ps was used.

\section{Acknowledgements} D. Bolmatov thanks Myerscough Bequest and K. Trachenko thanks EPSRC for financial support. D. B. acknowledges Thomas Young Centre for Junior Research Fellowship and Cornell University (Neil Ashcroft and Roald Hoffmann) for hospitality.

\section{Author contributions}
All authors have contributed equally to this work.

\section{Competing financial interests}
The authors declare no competing financial interests.

\end{document}